# Linear Power Grid State Estimation with Modeling Uncertainties

Martin R. Wagner[1], *Graduate Student Member*, Marko Jereminov[1], *Graduate Student Member*, and Larry Pileggi[1], *Fellow, IEEE*

*Abstract*—Recent advances in power system State Estimation (SE) have included equivalent circuit models for representing measurement data that allows incorporation of both PMU and RTU measurements within the state estimator. In this paper, we introduce a probabilistic framework with a new RTU model that renders the complete SE problem linear while not affecting its accuracy. It is demonstrated that the probabilistic state of a system can be efficiently and accurately estimated not only with the uncertainties from the measurement data, but also while including variations from transmission network models. To demonstrate accuracy and scalability we present probabilistic state estimation results for the 82k test case that represents the transmission level grid of the entire USA. It is shown that the estimated state distributions include the true grid state, while their mean exactly corresponds to the estimated deterministic state obtained from the nonlinear state estimator.

## I. INTRODUCTION

Reliable operation and control of an electrical power system is a very complex task, requiring a power system operator to perpetually monitor the current state of the system. This is accomplished by collecting measurement data from a multitude of points and variables in the system, each of which include some uncertainty attached to them. To account for the errors, uncertainties and other real-world problems such as faulty communication to the operator, the raw data is processed in a way to find the most likely state of the system. Hence, an accurate and robust State Estimator (SE) is an important part of this data processing tool chain to guarantee reliable power system operations decisions [1]. However, SE formulations are generally nonlinear and as such can suffer from non-convergence in real world scenarios [2].

To observe the state of power systems two types of measurement devices are deployed. Traditionally, and most commonly used are Remote Terminal Units (RTUs), which provide measurements of the phasor voltage and current magnitudes as well as angle difference between them. With the availability of geolocation systems, like GPS, that facilitate precise time coordination, it becomes possible to monitor power systems with much higher precision. For instance, Phasor Measurement Units (PMUs) use time-synchronization to precisely measure voltage and current phasors. However, even though this technology was first introduced in the early 1990s, there are still relatively few PMUs employed in today's power systems due to cost. Therefore, SE algorithms based on assumptions of significant data uncertainty remained.

The first mathematical formulation for power system state estimation was introduced by Schweppe in 1968 and later improved in [3]-[4]. It assumes a Gaussian distribution of measurement errors as well as known grid topology, and further aims to estimate the state of the power system by minimizing the mean square of the measurement errors (MSE). In this formulation, the state of the power system is described in terms of voltage magnitudes and phase angles at every bus of the system model whose governing equations are defined in terms of a power mismatch formulation.

Formulating the network constraints in rectangular coordinates using power [5] or current mismatch equations [6] has been explored as well. The latter has the advantage of rendering PMU measurement data linear, and hence the linear state estimation algorithms for systems observed fully with PMU measurements are proposed in [7]-[8]. This is, however, not a realistic scenario in present day systems, since only 1500 PMUs were employed on the entire continental US as recently as 2016 [9]. Therefore, different schemes have been proposed to combine hybrid measurement datasets within a state estimator. Namely, multistage algorithms handling RTU measurements and PMU data in a separate stages [10], as well as formulations integrating both kinds of measurements simultaneously [11]. Notably, the aforementioned SE algorithms use the assumption of independent measurements, while some work has been done to address this assumption and propose improved algorithms [12].

The equivalent circuit based formulation for power-flow simulations was recently introduced [13] as an extension of previous work on current-voltage formulations [14]. It was shown that any physics based data, including that from measurement devices such as PMUs and RTUs [15], can be represented by an equivalent circuit model within the proposed framework [13]. These circuit-based models and corresponding adaptation of circuit simulation algorithms were demonstrated to improve convergence robustness and scalability to large system sizes [13]. Additionally, the circuit formalism provides insight regarding the physical characteristics of the power grid, thereby enabling new modeling approaches [16].

[†]This work was supported in part by the Defense Advanced Research Projects Agency (DARPA) under award no. FA8750-17-1-0059 for the RADICS program, and the National Science Foundation (NSF) under contract no. ECCS-1800812.

[1]Authors are with the Department of Electrical and Computer Engineering, Carnegie Mellon University, Pittsburgh, PA 15213 USA (e-mail: {mwagner1, mjeremin, pileggi}@andrew.cmu.edu).

The equivalent circuit formulation was also demonstrated to enable optimization within the same framework, referred to as Equivalent Circuit Programming (ECP). This formulation was applied to find power flow feasibility in [17] and optimal power flow solution in [18]. In an ECP formulation, the objective function together with its constraints is modeled as a circuit problem, where the operating point of the circuit represents a solution of the optimization problem. As with other equivalent circuit formulations, it is possible to apply circuit simulation heuristics to ensure convergence for what are otherwise very difficult nonlinear problems. A linear state estimation algorithm, based on an equivalent circuit formulation, and including RTUs and PMUs, was demonstrated in [19]. In this paper we rederive a linear RTU model that is enabled by the feasibility power flow capability in [17], and use it to formulate an ECP problem that represents a linear state estimation that includes both RTU and PMU data. We show that the introduced formulation represents a deterministic linear SE algorithm that does not sacrifice accuracy. Additionally, since no iterative process is required to solve for the estimated state, this algorithm does not suffer from convergence issues for real world systems as nonlinear algorithms do [2]. Given the efficiency of the problem solution, we can further evaluate the probabilistic state using Monte Carlo analyses. Our linear algorithm is compared to a nonlinear algorithm using a recently proposed RTU model [15]. Furthermore, we evaluate the influence of less reliable data and propose a weighting scheme for both the linear and nonlinear algorithms to remedy the effects of such data. To demonstrate the scalability of the linear probabilistic algorithm, results are presented for an 82k-bus case that represents a transmission-level power grid of the United States [20].

## II. BACKGROUND

### A. Equivalent Circuit Formulation (ECF)

The equivalent circuit formulation was shown to represent a robust and scalable framework for simulating the steady-state response of a power system [13]. As with the current injection method, the governing power flow equations are formulated in terms of current and voltage state variables. This eliminates the inherent nonlinearities introduced by a power mismatch formulation in modeling the transmission network constraints. Of course, the nonlinearities are now introduced in representing the constant power generator and load models, thereby resulting in a different set of nonlinear equations.

Each of the power flow models is represented by an equivalent circuit to facilitate the use of the circuit formalism in solving the nonlinear equations as follows. First, they are defined in terms of phasor current and voltage state variables. To provide the analyticity of models that constrain constant power, the nonlinear set of complex KCL equations is split into real and imaginary parts. The resulting nonlinear models now represent real valued functions that are further linearized to facilitate the use of Newton-Raphson (N-R). The linearized and split equations are then mapped into linear coupled real and imaginary equivalent circuits, as shown for the generator in the 3-bus example in Fig. 1. These split equivalent circuits are the linearized sub-problems that are iteratively solved and updated using the N-R algorithm to find a solution to the original nonlinear problem. This formalism is identical to the solution formalism for nonlinear circuits in the circuit simulation field. Most importantly, it provides a new perspective on the power flow problem and creates the opportunity to adapt algorithms from the circuit simulation domain into the power systems domain. More details of this approach and the corresponding results for large, real-world systems can be found in [13],[21].

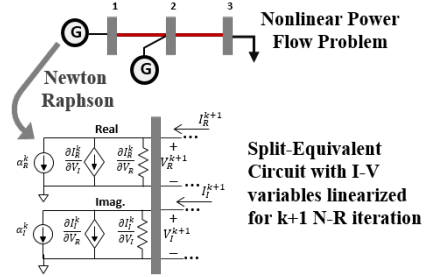

*Fig. 1 Each system component (generator in this figure) is mapped to an equivalent split-circuit model in terms of impedances, controlled sources and independent sources.*

### B. Equivalent Circuit Programming

An extension of the ECF was the development of an Equivalent Circuit Programming (ECP) formulation that was described in [17]-[18] to solve optimization problems by formulating them as equivalent circuits; namely, by representing them as equivalent circuits while following mathematically well-defined optimization formalisms. Many optimization problems in the power system domain can be tackled that way. A first example is presented in the formulation for feasibility analysis of power flow problems shown in [17].

The foundation of ECP is laid by the concept of adjoint networks that was introduced for circuit design and analysis in the 1960s [22][23]. Adjoint networks capture sensitivities of the original network, and can be used for sensitivity analysis [22] or circuit optimization [23]. Formulations based on the adjoint network theory were also proposed for power system analysis in [24],[25].

Adjoint circuits are derived from Tellegen's theorem, which states that for branch current and voltage row vectors $I, V$ in a given primal network $N$ and its adjoint branch current and voltage row vectors $\mathfrak{I}, \lambda$ in a topologically equivalent adjoint network $\hat{N}$ the following holds

$$I^T V = \mathfrak{I}^T \lambda = \mathfrak{I}^T V = I^T \lambda = 0 \quad (1)$$

$$I^T \lambda - \mathfrak{I}^T V = 0 \quad (2)$$

Hence, (2) is true for $N$ and $\hat{N}$ as well. A linear circuit with a complex admittance matrix $Y$ can be formulated as

$$I = YV. \quad (3)$$

Combining (2) and (3) results in

$$V^T (Y^H \lambda) = V^T \mathfrak{I} \quad (4)$$

which in order to satisfy Tellegen's theorem defines the admittance matrix of the adjoint circuit to be

$$Y_{adj} = Y^H. \quad (5)$$

This defines a mapping for linear circuit elements in the original network to linear elements of its adjoint network. As described in [22], the mapping from original to adjoint circuit sets all primal sources to zero, creating a trivial operating point for the adjoint circuit. A formalism to define the adjoint elements of nonlinear circuit elements is discussed in [17]-[18], and follows the same steps as described in (1)-(5).

With the defined relationship between the primal and adjoint circuit, the governing equations of an ECP problem are obtained by coupling the primal and its adjoint (dual) circuit, which corresponds to obtaining the necessary KKT optimality conditions [26] of an optimization problem. In the case of a power grid optimization problem, the ECP governing equations represent the real and imaginary equations of the power flow formulation (ECF) that are coupled with the respective real and imaginary parts of the adjoint network equations. These adjoint network equations also represent the dual equations from the necessary KKT conditions [26], and apart from two added terms, have the same structure as the adjoint networks defined by Tellegen's theorem. First, for the formulation proposed here, gradients of the optimization's objective function are added as adjoint sources [18]. Secondly, terms that define the coupling between adjoint and primal networks are added that also augment the primal network equations [17]. A coarse explanation of the mechanism of the resulting optimization algorithm is that the coupling between the adjoint and the primal networks steers the original system to an optimal solution in terms of the objective function that is embedded in the adjoint network equations. This description holds for optimizations that include equality constraints. To incorporate inequality constraints, refer to [18] for the corresponding circuit-based modeling and derivation of the complementary slackness conditions. Lastly, by looking at these equations as equivalent circuits, circuit simulation methods can be exploited to develop efficient heuristics to solve these otherwise hard to solve problems.

*C. Feasibility Sources as optimization tool*

ECP can be used to formulate an optimization problem to determine and quantify infeasibilities in power flow simulations [17]. This capability is critical for enabling the derivation of the linear RTU model proposed in this paper, therefore it is briefly described here.

The Feasibility Power Flow algorithm is derived as follows. A solution to a power flow problem can be found by linearizing and iteratively solving a set of current injection equations:

$$Y_{GB}V + I(V) = 0 \quad (6)$$

where $Y_{GB}$ represents the split-admittance matrix, while $I(V)$ represents the nonlinear current injection vector (e.g. constant power models). Importantly, due to the nonlinear currents $I(V)$, the solution to (6) may not be able to satisfy KCL at every node, thereby resulting in an infeasible power flow case. With the feasibility optimization approach, current sources, $I_F$, are added to all (or selected) nodes of a system, and further minimized to locate and quantify the power flow infeasibilities. Namely, $I_F$ is minimized to zero if the power flow solution exists. Nonzero $I_F$ values indicate an infeasible power flow problem, and their magnitudes represent the minimum current that is required (at those specific nonzero locations) to obtain a feasible solution. Mathematically this optimization problem is defined as:

$$\min_{I_F} \frac{1}{2}\|I_F\|_2^2 \quad (7)$$
$$\text{s. t.} \quad Y_{GB}V + I(V) = I_F \quad (8)$$

A solution to the constrained optimization problem from (7)-(8), is found by formulating the Lagrangian as

$$\mathcal{L}(V, I_F, \lambda) = \frac{1}{2}\|I_F\|_2^2 + \lambda^T(Y_{GB}V + I(V) - I_F) \quad (9)$$

and differentiating it to obtain the necessary optimality (KKT) conditions:

$$\frac{\partial \mathcal{L}}{\partial \lambda} \to Y_{GB}V + I(V) - I_F = 0 \quad (10)$$

$$\frac{\partial \mathcal{L}}{\partial V} \to [Y_{GB}^T + \mathcal{J}^T(V)]\lambda = 0 \quad (11)$$

$$\frac{\partial \mathcal{L}}{\partial I_F} \to I_F = \lambda \quad (12)$$

Where (10) are the power flow equations with included feasibility sources representing the primal problem (as split-circuits), while (11) defines the governing adjoint circuit equations, with the first order sensitivity matrix $\mathcal{J}(V)$, as derived in [17]. Next, (12) provides the relationship between the feasibility currents $I_F$ and the Lagrange multipliers $\lambda$ that can be used to further eliminate the $I_F$ variables. As described in Section II. B., the primal and dual problems from (10)-(11) represent the governing equations of two coupled nonlinear split-circuits (primal and adjoint) that can be linearized and iteratively solved by applying circuit simulation techniques [21]. Furthermore, if a calculated operating point satisfies the second order sufficiency KKT conditions [26], it also represents an optimal solution to the power flow feasibility problem [17].

### III. EQUIVALENT CIRCUIT MODELS OF MEASURED DATA AND STATE ESTIMATION PROBLEM

Power System State Estimation (SE) can be formulated as an optimization problem that is similar to the aforementioned feasibility power flow formulation. Namely, it is equivalent to finding the closest feasible system state given the network constraints with incorporation of PMU and RTU measurement devices.

*A. Network Model*

For SE algorithms, the network topology that is represented by transmission lines, transformers, and shunts, is assumed to be known. Our approach is based on models that are identical to those used for power flow and other SE algorithms [13]. Since network equations are linear in current-voltage based formulations [27], the network equations governing the SE algorithm are linear as well [13].

*B. PMU Model*

A PMU measurement model consists of real and imaginary voltage measurements on a bus, as well as real and imaginary current measurements for branches from that bus. To make use of both types of measurements they are combined in the PMU model with current source conductances, as seen in Fig. 2.

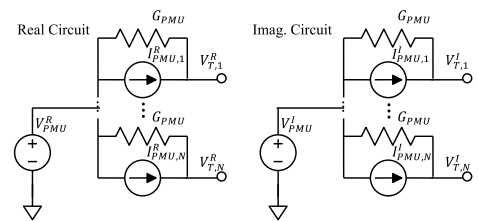

*Fig. 2 Split Circuit Model of PMU measurements.*

For each branch measurement the current $I_G$ through the source conductance $G_{PMU}$ is a metric for the measurement error of the PMU. For example, if the measurement values $V_{PMU}^{R,I}$ and $I_{PMU}^{R,I}$ perfectly depict the state of the system, the currents $I_{PMU}^{R,I}$ inject just as much current into the system that the terminal voltages $V_T^{R,I}$ are equal to the voltages $V_{PMU}^{R,I}$, thereby setting the current through the source conductances $I_{G_{PMU}}^{R,I}$ to zero. A deviation of any of the measured values ($V_{PMU}^{R,I}$, $I_{PMU}^{R,I}$) from the real state of the system causes a current proportional to the resulting voltage difference to flow through the respective conductances $G_{PMU}$. Hence, an algorithm to minimize measurement errors can be designed by minimizing the currents through the PMU source conductances $G_{PMU}$ ($I_{G_{PMU}}$).

### C. RTU Models

Assuming that an RTU measures the voltage magnitude on a bus $V_M$ and the apparent power on a branch from that bus ($S_{RTU} = P_{RTU} + j Q_{RTU}$), we can define an admittance $Y_m$ that maps to the mean measurement values [15],

$$Y_m = G_m + jB_m = \frac{S_{RTU}}{V_M^2} \quad (13)$$

and preserves angle as well as amplitude information of the original measurements. Each value for $G_m$ and $B_m$ can be negative, depicting generation of real power, and capacitive or inductive behavior of the bus.

#### 1) Nonlinear $\Delta Y_{RTU}$-RTU model

To derive the nonlinear $\Delta Y_{RTU}$-RTU model [15], we first define the difference between the RTU measurement means ($G_m, B_m$) and the variables of the RTU model ($G_{RTU}, B_{RTU}$) as:

$$\Delta G = G_{RTU} - G_m, \quad (14)$$
$$\Delta B = B_{RTU} - B_m. \quad (15)$$

Therefore, minimizing the $\Delta Y_{RTU}$-RTU measurement error corresponds to minimizing the conductance/susceptance mismatch, $\Delta G$ and $\Delta B$. Most importantly, since the RTU model represents a shunt element, its governing equations can be defined in terms of real and imaginary RTU current injections [15]:

$$I_R^{\Delta Y} = (\Delta G + G_m)V_R + (\Delta B + B_m)V_I, \quad (16)$$
$$I_I^{\Delta Y} = (\Delta G + G_m)V_I - (\Delta B + B_m)V_R, \quad (17)$$

where $G_{RTU}$ and $B_{RTU}$ are now expressed in terms of (14)-(15). The circuit representation of (16) and (17) is shown in Fig. 3. Importantly, the current terms proportional to RTU conductance/susceptance difference variables are nonlinear, hence the complete SE problem becomes nonlinear [15].

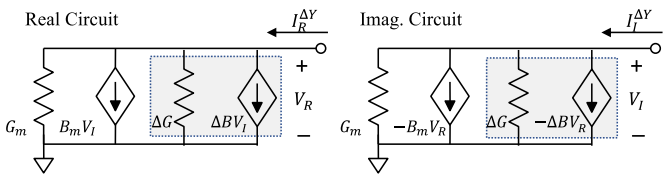

*Fig. 3 Equivalent circuit representation of a $\Delta Y_{RTU}$-RTU model.*

#### 2) Linear $\Delta I_{RTU}$-RTU Model

A linear RTU equivalent circuit model was recently proposed in [19], where RTU measurements are represented by current sources. Therein, the RTU measurement error is optimized by minimizing the difference between the RTU current variables and terms equivalent to (16)-(17) at the measurement means. In this paper, we use the circuit formalism to further understand and rederive the linear RTU [19] and present the $\Delta I_{RTU}$-RTU model defined in terms of the recently introduced feasibility currents [17].

To arrive at the $\Delta I_{RTU}$-RTU model, we replace the $\Delta G$ and $\Delta B$ terms that introduce the nonlinearities within the $\Delta Y_{RTU}$-RTU with feasibility current sources [17]. Now, instead of minimizing the conductance/susceptance mismatch to obtain the estimated system state, the current mismatches $\Delta I_{RTU}^{R,I}$ are minimized.

$$I_R^{\Delta I} = G_m V_R + B_m V_I + \Delta I_{RTU}^R, \quad (18)$$
$$I_I^{\Delta I} = G_m V_I - B_m V_R + \Delta I_{RTU}^I, \quad (19)$$

The equivalent circuit model of the $\Delta I_{RTU}$-RTU is depicted in Fig. 4.

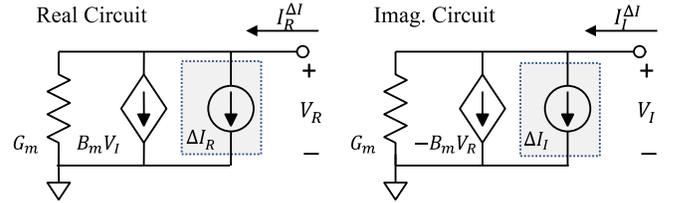

*Fig. 4 Equivalent circuit of the $\Delta I_{RTU}$-RTU.*

As we will see in the results section, the $\Delta I_{RTU}$-RTU and $\Delta Y_{RTU}$-RTU concepts provide very similar results. The main advantage of the $\Delta I_{RTU}$-RTU model lies in the fact that only linear circuit elements are used, and hence, given linear network constraints and a linear PMU model, the whole state estimation algorithm can be expressed as a *linear problem*.

### D. Defining the SE optimization problem

We can now define different deterministic SE algorithms by combining any of the introduced RTU and/or PMU models with the models of a power system topology as optimization problems that minimize the differences from the measurement means. This formulation is equivalent to minimizing the squares of the measurement errors under the constraint of fulfilling the network equations. The objective function of an algorithm including PMU models and one of the two RTU models can be defined as:

$$\min_X \mathcal{F}_e = \left\|I_{G_{PMU}}^R\right\|_2^2 + \gamma_{RTU} \begin{cases} \Delta Y_{RTU}: & \|\Delta G\|_2^2 + \|\Delta B\|_2^2 \\ \Delta I_{RTU}: & \|\Delta I_{RTU}^R\|_2^2 + \|\Delta I_{RTU}^I\|_2^2 \end{cases} \quad (20)$$

that is subjected to network constraints. Herein, $X$ is the state vector of the optimization problem, $I_{G_{PMU}}^{R,I}$ represents the vector of real and imaginary currents that correspond to the mismatch in the PMU models, and depending on the choice of the RTU model, the objective is either to minimize their conductance/susceptance mismatch or their infeasibility currents.

Importantly, the choice of the $\Delta Y_{RTU}$-RTU model results in a nonlinear algorithm, while by choosing the $\Delta I_{RTU}$-RTU model, a fully linear deterministic SE algorithm is defined.

To augment these formulations, the vector of RTU weighting factors, $\gamma_{RTU}$, can be used to account for RTU data with variable uncertainty attached to it. Weighting of measurements based on their uncertainty is proposed in the classical SE algorithm [3], and provides an important addition to account for the real world SE challenge of unreliable data. It is important to note that, by design, our PMU models include a weighting factor implicitly within the equivalent circuit representation. The conductance $G_{PMU}$ serves as a weighting factor by adjusting the currents $I_{G_{PMU}}^{R,I}$ in relation to the difference between the measurement voltage and the terminal voltage.

While equivalent circuit models for SE were used in [15], and the defined nonlinear optimization problem was solved within a commercial optimization toolbox, for this work we derive models in terms of ECP and natively solve the optimization using a C++ implementation of the equivalent circuit based simulation tool, SUGAR (Simulation of Unified Grid Analysis and Renewables) [13].

### E. ECP for solving the SE Problem

To find the full PMU model for ECP, we continue by deriving its adjoint circuit model by shorting the voltage sources of Fig. 2 and opening the corresponding current sources [17]. This results in the linear adjoint circuit that consists of a conductance $G_{PMU}$ connected to ground.

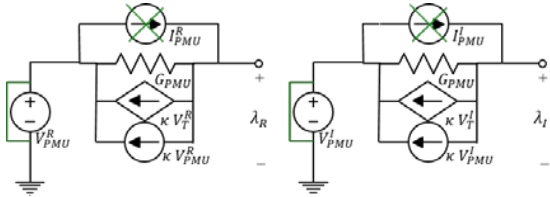

Fig. 5 Adjoint equivalent circuit of the PMU model. Here, sources of the primary model are off, $V_T^{R,I}$ are variables (the primary voltages), $V_{PMU}^{R,I}$ are constants (the measurement values), and $\kappa$ is $2\,G_{PMU}^2$.

Furthermore, the gradient information of the objective function that augments the adjoint circuit is derived by expressing the real and the imaginary $I_{G_{PMU}}^{R,I}$ as

$$I_{G_{PMU}}^{R,I} = G_{PMU}\big(V_T^{R,I} - V_{PMU}^{R,I}\big). \tag{21}$$

Plugging (21) into (20) and calculating the gradient

$$-\nabla\,\mathcal{F}_e(X) = 2\,G_{PMU}^2\,V_T^{R,I} - 2\,G_{PMU}^2\,V_{PMU}^{R,I}, \tag{22}$$

corresponds to the values of the additional adjoint sources. Here, the terms proportional to the $V_{PMU}^{R,I}$ (the PMU voltage measurement values) are modeled as constant current sources and the terms proportional to $V_T^{R,I}$ (PMU terminal voltage variables) are modeled as controlled sources of the corresponding voltages, $V_T^{R,I}$. This leads to the adjoint part of a PMU's ECP model that is depicted in Fig. 5.

The full RTU models are derived in a similar fashion. For both models, their adjoint admittance provides the basis of their ECP models. Now, the construction of the full $\Delta I_{RTU}$-RTU model is straight forward, since an infeasibility current source does not have an adjoint circuit element, but rather is directly coupled to the primal circuits by the adjoint voltages, $\lambda^{R,I}$ [17]. To derive the final part of the adjoint ECP model of a $\Delta Y_{RTU}$-RTU, the formalism to derive nonlinear adjoint models as described in [17] is used. In addition to the linear adjoint admittance, this results in bilinear equations in terms of the models' variables ($G_{RTU}, B_{RTU}$) with $V_{R,I}$, as well as with $\lambda_{R,I}$, that are further linearized to enable solution via N-R. The use of the nonlinear RTU model results in a nonlinear optimization algorithm. Hence, numerically obtained solutions are verified by formulating sufficient conditions to ensure convergence to an optimum [26].

### F. Probabilistic State Estimation

Historically, due to the computational complexity and unreliable convergence properties, state estimation algorithms were formulated in a deterministic way to find the most likely power system state under measurement uncertainties. However, measurement uncertainties are not the only uncertain component in the problem. Parameters of network models are subject to uncertainty as well. One reason for this is that they are usually modeled for specific environmental parameters. A probabilistic picture of the estimated system state including measurement uncertainties and network uncertainties enables a much better understanding of possible grid states for the investigated set of measurements. For example, with a probabilistic algorithm it is now possible to quantify risks in the power system state in a probabilistic way, providing additional information for early decision making.

In this paper we use a Simple Random Sampling Monte Carlo simulation approach on top of the linear SE algorithm to arrive at a linear probabilistic state estimation algorithm. This SE algorithm builds on a Monte Carlo simulations approach for probabilistic power flow that was introduced in [28] and uses a thread level parallel extension of SUGAR in C++. This approach was augmented to include a modern, fast pseudo random number generator with multiple parallel streams and very good statistical properties [29]. A separate data pass was added to effectively adapt the linear probabilistic parameters of the network, PMU, and RTU models for each Monte Carlo sample. The results show that the efficiency of this formulation enables scaling to very large system sizes.

## IV. RESULTS

Our synthetic test cases were generated as follows. State estimation sample cases were first created by solving a power flow test case within the SUGAR C++ code base [28]. This power flow solution is interpreted as the true system state, and the vector of complex voltages is denoted as $\bar{X}$. Then, based on statistical inputs, PMU and RTU measurement models were randomly assigned to every bus of the system, thereby replacing the original power flow models on that bus. The power flow topology remains unchanged. The final step to create a deterministic SE case is to assign PMU and RTU measurement values by superimposing randomly created measurement errors on the pre-calculated true system states. All of this is done within the same code base to enable effective creation of multiple deterministic SE samples to calculate statistics for the performance measures of the algorithms that will be compared. The assignment of a PMU or RTU model to a certain bus remains unchanged when multiple deterministic SE samples are created.

In the following experiments the accuracy of the state estimation algorithms is measured by the sum of squared errors over the state estimation solution vector

$$\sigma_{ss} = \big(\hat{X}_{est} - \bar{X}\big)^T\big(\hat{X}_{est} - \bar{X}\big), \tag{23}$$

as well as the maximum error of a solution component

$$\sigma_{max} = \max|\hat{X} - \bar{X}|. \tag{24}$$

Both measures are calculated from the voltage solution vector in rectangular coordinates.

The deterministic algorithms are tested for five power system cases, ranging from 500 to 82k buses. They represent a synthetic power grid model of the state of South Carolina with 500 buses (SC) [20], a model of the French transmission system with 6515 buses (RTE) [30], a model of a pan European transmission system with 13659 buses (PEGASE) [30], as well as synthetic models of the US eastern interconnection with about 70k buses (East), and a synthetic 82k bus system model of the entire USA [20].

For the presented experiments we assume that 4% of system buses are observed by PMU measurement data without error and 6% of the system buses have PMU measurements including a measurement error. All other buses are observed by RTUs.

All PMUs use the same model with a source conductance $G_{PMU}$ of 10 p.u. Imperfect PMUs are assumed to have normally distributed errors with a standard deviation of 0.02% of the mean of each measured value ($V_{PMU}^{R,I}, I_{PMU}^{R,I}$).

### A. Comparing the $\Delta I_{RTU}$-RTU to the $\Delta Y_{RTU}$-RTU model

In the first experiment, the linear $\Delta I_{RTU}$-RTU is compared to the nonlinear $\Delta Y_{RTU}$-RTU model. Here, all RTUs are assumed to have a normally distributed voltage magnitude measurement error with a standard deviation of 0.4% and a normally distributed error with a standard deviation of 1% for the measurements of real and reactive power. Since all RTUs are modeled as having the same uncertainty, RTU weighting factors are not used in this first experiment in Table 1.

Table 1 Comparison between linear $\Delta I_{RTU}$-RTU and nonlinear $\Delta Y_{RTU}$-RTU. To arrive at measures of $\sigma_{SS}$ and $\sigma_{max}$ state estimation cases are created until the 99% confidence interval is smaller than 5% of the mean of the measures.

| $\sigma_{SS}$ | $\Delta I_{RTU}$-RTU | $\Delta Y_{RTU}$-RTU |
|---|---|---|
| SC | 1.24e-4 | 9.29e-5 |
| RTE | 3.13e-3 | 2.62e-3 |
| PEGASE | 1.01e-2 | 9.39e-3 |
| East | 9.45e-3 | 1.16e-2 |
| USA | 1.13e-2 | 1.28e-2 |
| $\sigma_{max}$ | $\Delta I_{RTU}$-RTU | $\Delta Y_{RTU}$-RTU |
| SC | 2.92e-3 | 2.80e-3 |
| RTE | 7.14e-3 | 6.81e-3 |
| PEGASE | 1.25e-2 | 1.42e-2 |
| East | 3.24e-2 | 4.68e-2 |
| USA | 2.82e-2 | 4.04e-2 |

Since SE cases are created by a randomized algorithm, multiple state estimation cases have to be evaluated to make valid statements about an algorithm's quality. In this paper we evaluate deterministic SE samples until the remaining uncertainty of the created measure (i.e. $\sigma_{SS}$ or $\sigma_{max}$) is less than 5% of its mean value in a 99% confidence interval. For example, for the $\Delta I_{RTU}$-RTU algorithm applied to the SC test case, 578 SE cases were evaluated until the 99% confidence interval for the $\sigma_{SS}$ measure was smaller than [1.24E-4 ± 1.24E-4*0.05].

With this in mind, we can state that the $\Delta I_{RTU}$-RTU results are similar to the results of the $\Delta Y_{RTU}$-RTU. In fact, for bigger systems, the linear $\Delta I_{RTU}$-RTU outperformed the $\Delta Y_{RTU}$-RTU model while providing a clear computational advantage.

This is further demonstrated by the result in Fig. 6 that compares the SE error distributions based on both models for the USA testcase. For this testcase, the majority of the SE error stems from a small number of buses, thus explaining the similar results for the SE performance measures, $\sigma_{SS}$ and $\sigma_{max}$.

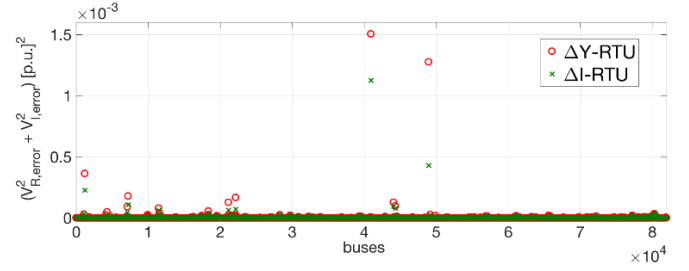

Fig. 6 Sum of squared voltage errors ($V_{R,error}^2 + V_{I,error}^2$) for each bus of the 82k bus USA system based on the two different RTU models for the same state estimation case.

### B. Including RTU weighting factors

To study the influence of RTU weighting factors on the SE algorithms we assume that for 10% of the RTU measurements no current data is available. This accounts for various situations, such as unreliable communication to the affected measurement devices. To make a good guess for the missing data, we assume that under normal conditions the system state changes little between intervals and estimate the measurement as being the value from the previous cycle. For these measurements we assume a 10 times higher standard deviation, resulting in a 10 times lower weight for the affected measurements. For all other measurements we keep the same statistical assumptions as specified in the first experiment.

Table 2 Comparison of $\sigma_{SS}$ and $\sigma_{max}$ measures for weighted and non-weighted RTUs and for systems with 10% of RTUs having ten times higher uncertainty.

| | $\Delta I_{RTU}$-RTU | | $\Delta Y_{RTU}$-RTU | |
|---|---|---|---|---|
| $\sigma_{SS}$ | weighted | unweighted | weighted | unweighted |
| SC | 7.82e-4 | 1.01e-3 | 4.19e-4 | 7.63e-4 |
| RTE | 4.54e-2 | 7.91e-2 | 2.90e-2 | 7.21e-2 |
| PEGASE | 2.10e-1 | 2.45e-1 | 1.49e-1 | 2.17e-1 |
| East | 1.16e-1 | 1.70e-1 | 5.71e-2 | 1.82e-1 |
| USA | 1.84e-1 | 2.47e-1 | 9.32e-2 | 2.41e-1 |
| $\sigma_{max}$ | weighted | unweighted | weighted | unweighted |
| SC | 5.56e-3 | 7.72e-3 | 4.82e-3 | 7.39e-3 |
| RTE | 4.88e-2 | 7.26e-2 | 4.29e-2 | 7.10e-2 |
| PEGASE | 6.05e-2 | 6.66e-2 | 5.57e-2 | 6.53e-2 |
| East | 4.82e-2 | 8.35e-2 | 3.94e-2 | 1.42e-1 |
| USA | 5.57e-2 | 9.77e-2 | 4.72e-2 | 1.21e-1 |

Table 2 shows the influence of RTU weighting factors for the measures $\sigma_{SS}$ and $\sigma_{max}$. These results are created using the same algorithm that was applied in the first experiment to arrive at 99% confidence intervals that are no bigger than 5% of the corresponding mean.

Creating deterministic SE samples where 10% of the measurements have an order of magnitude range of standard deviation values is testing unweighted algorithms to the extreme. Nonetheless, both unweighted algorithms are able to recover high quality SE results for all cases. Just as in the first experiment, the $\Delta I_{RTU}$-RTU performs similar to the $\Delta Y_{RTU}$-RTU and keeps outperforming the nonlinear algorithm for bigger cases.

After including RTU weighting, the results show that both algorithms recover an improved state estimate for every system. However, RTU weighting improves the SE result of the nonlinear algorithm more than the result of the linear algorithm.

Now the $\Delta Y_{RTU}$-RTU slightly outperforms the $\Delta I_{RTU}$-RTU for all cases.

## C. Probabilistic State Estimation

Results for the linear probabilistic state estimation algorithm are presented for a synthetic USA transmission power system case [20] with 82k buses. As in the previous examples, we assume that 10% of system buses are observed by PMUs, such that, 40% are perfect measurements and 60% have 0.02% normally distributed measurement uncertainty ($V_{PMU}^{R,I}$, $I_{PMU}^{R,I}$). Furthermore, RTUs have 1% normally distributed measurement uncertainty for their real and reactive power injections, and 0.4% for their voltage magnitude measurement.

With this, a deterministic SE case based on the synthetic USA system [20] is created and evaluated with the linear probabilistic SE algorithm. For each of the following experiments, $10^4$ Monte Carlo samples are created by sampling from the uncertainty distributions of the SE case.

While both, the deterministic SE case, and the Monte Carlo sample are created by a randomized algorithm using the same uncertainty distributions, a clear distinction has to be made between them. The deterministic SE case is created by adding randomized errors to the previously calculated true system states, while a Monte Carlo sample is created without the knowledge of the true system state by reintroducing uncertainty distributions around the randomized measurements.

In the first probabilistic experiment, only measurement uncertainties of PMUs and RTUs are included. Results for four selected buses are presented as green probability density functions (PDFs) of voltage magnitudes in Fig. 7 and of voltage angles in Fig. 8.

The distributions change notably with the distance of the observed bus from the closest PMU measurement. For Fig. 7 and Fig. 8, the closest PMU-measured buses are: (a) 4 hops away, (b) 5 hops away, (c) one hop away, and (d) is a PMU measured bus. The difference in the SE comparisons between (c) and (d) in Fig. 7 and Fig. 8, however, is not due to the one hop difference in distance from the closest PMU measurement bus, but due to the difference between imperfect and perfect PMU measurements.

While there is a clear advantage in knowing the probabilistic picture of a state estimation result, the full potential of this algorithm is only applied if all uncertainties in the model are quantifiable in a realistic way. By including all known uncertainties and possibly their correlations, more meaningful probabilistic results are created. To explore the impact of this additional uncertainty, we add models for transmission line and transformer variations. To quantify these uncertainties, we include a temperature dependence on resistive losses that adds to the base uncertainty of the modeled components, resulting in the assumed uncertainties for the second probabilistic experiment presented in Table 3.

*Table 3 Standard deviations of normally distributed uncertainty values of transmission line and transformer series elements.*

| Network uncertainties | $\sigma_R$ [% of mean] | $\sigma_X$ [% of mean] |
|---|---|---|
| Transmission line | 5% | 0.5% |
| Transformer | 0.5% | 0.1% |

We compare the probabilistic results of the second experiment that are presented as black PDFs in Fig. 7 and Fig. 8, with the ones from the first experiment. The additional uncertainties produce a widening of the PDFs for some buses, establishing a non-trivial influence of the network uncertainties on the probabilistic states. For example, in Fig. 7, figures (a) and (d) show close to zero difference in their standard deviations while pictures (b) and (c) display a clear widening in their possible states. Little influence of network uncertainty is not surprising for the PMU measured bus (d), but not as simple to explain for the RTU measured bus (a).

By comparing figures (b) and (c) of Fig. 7 and Fig. 8, we find that the effects of the network uncertainties on the voltage angles are less than those on voltage magnitudes. This is explained by the relatively higher dependence of the voltage magnitudes on the resistive losses.

The ability to observe and quantify the effects of model uncertainties on different system variables clearly present the advantages of probabilistic algorithms for SE.

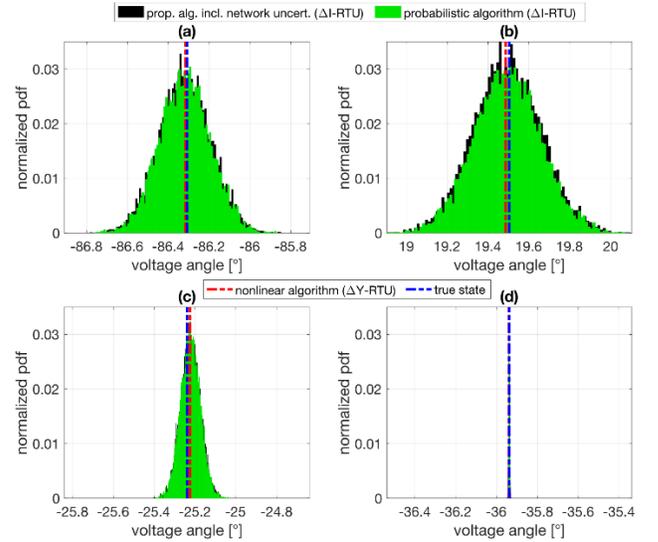

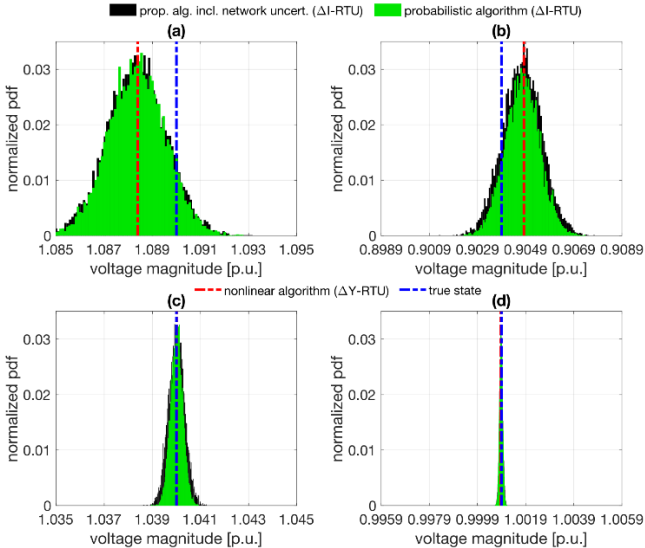

*Fig. 7 Selected voltage magnitude pdfs of the probabilistic SE results for the 82k+ synthetic USA case.*

*Fig. 8 Selected voltage angle pdfs of the probabilistic state estimation results for the synthetic USA test case.*

Notably, the true state of the (synthetic data) system, plotted in a dashed vertical blue line in Fig. 7 and Fig. 8, is always inside the PDFs of the probabilistic result. We also find that the mean of the probabilistic ($\Delta I_{RTU}$-RTU) results almost perfectly coincides with the solution of the nonlinear deterministic $\Delta Y_{RTU}$-RTU algorithm, plotted in a dashed vertical red line in Fig. 7 and Fig. 8.

*D. Discussion of the probabilistic algorithm*

The linear probabilistic algorithm that was implemented in C++ was used to evaluate $10^4$ Monte Carlo samples of the 82k bus case in 23 minutes on an Intel Xeon server CPU (E5-2620) running on 2.10GHz using 28 of 32 possible threads. Monte Carlo samples of this algorithm are independent and have close to constant runtime, hence it is expected to scale linearly with additional CPUs. Taking this into account, we think that the probabilistic algorithm has the potential to scale well enough to be used in an operations setting for interconnection size systems with currently available hardware, where off the shelf shared memory servers can scale to hundreds of CPU cores.

The current algorithm assumes independence of every probabilistic variable, which is generally not the case. Measurements that are only connected through a single network connection, or more general measurements that are in close proximity to each other, will have some correlation with each other. Future work will consider adding models of these correlations to further improve the quality of the results.

## V. Conclusion

This paper describes a novel linear probabilistic state estimation algorithm based on formulating the SE as optimization problem that is expressed in terms of equivalent circuit models. A PMU model and a nonlinear RTU model were re-derived and presented for this formulation. A linear RTU model was further proposed to facilitate a fully linear state estimation algorithm. Both RTU models were compared and evaluated using measures of state estimation quality. Both models were found to produce comparable results for the studied systems. For bigger systems, the linear RTU model was found to outperform the nonlinear model. To address data with different uncertainty, weighted RTU algorithms were explored. Finally, a linear probabilistic state estimation algorithm including network uncertainties was proposed and shown to scale to large system sizes that are representative of entire transmission systems for large countries, such as the US.

## VI. References

[1] A. Monticelli, "Electric power system state estimation," *Proc. IEEE*, vol. 88, no. 2, pp. 262–282, 2000.
[2] T. D. Mohanadhas, N. D. R. Sarma, and T. Mortensen, "State Estimation Performance Monitoring at ERCOT," in *National Power Systems Conference (NPSC)*, 2016.
[3] F. C. Schweppe, "Power System Static-State Estimation, Parts I, II and III," no. 1, pp. 130–135, 1970.
[4] F. C. Schweppe and E. J. Handschin, "Static State Estimation in Electric Power Systems," *Proc. IEEE*, vol. 62, no. 7, pp. 972–982, 1974.
[5] N. D. Rao and L. Roy, "A Cartesian Coordinate Algorithm for Power System State Estimation," *IEEE Trans. Power Appar. Syst.*, no. 5, pp. 1070–1082, 1983.
[6] M. Lavorato, M. J. Rider, and A. V. Garcia, "Power system state estimation: A new method based on current equations," *LESCOPE'07 - 2007 Large Eng. Syst. Conf. Power Eng.*, pp. 166–170, 2007.
[7] L. Zhang, A. Bose, A. Jampala, V. Madani, and J. Giri, "Design, Testing, and Implementation of a Linear State Estimator in a Real Power System," *IEEE Trans. Smart Grid*, vol. 8, no. 4, pp. 1782–1789, 2017.
[8] D. A. Haughton and Heydt, "A Linear State Estimation Formulation for Smart Distribution Systems," *IEEE Trans. Power Syst.*, vol. 28, no. 2, pp. 1187–1195, 2013.
[9] U.S Department of Energy, "Advancement of synchrophasor technology in ARRA Projects," 2016.
[10] M. Zhou, V. A. Centeno, J. S. Thorp, and A. G. Phadke, "An alternative for including phasor measurements in state estimators," *IEEE Trans. Power Syst.*, vol. 21, no. 4, pp. 1930–1937, 2006.
[11] T. S. Bi, X. H. Qin, and Q. X. Yang, "A novel hybrid state estimator for including synchronized phasor measurements," *Electr. Power Syst. Res.*, vol. 78, no. 8, pp. 1343–1352, 2008.
[12] E. Caro, A. J. Conejo, and R. Mínguez, "Power system state estimation considering measurement dependencies," *IEEE Trans. Power Syst.*, vol. 24, no. 4, pp. 1875–1885, 2009.
[13] A. Pandey, M. Jerminov, M. R. Wagner, D. M. Bromberg, G. Hug, and L. Pileggi, "Robust Power Flow and Three Phase Power Analyses," *IEEE Trans. Power Syst.*, vol. 8950, no. c, pp. 1–10, 2018.
[14] P. A. N. Garcia, J. L. R. Pereira, S. Carneiro, M. P. Vinagre, and F. V. Gomes, "Improvements in the representation of PV buses on three-phase distribution power flow," *IEEE Trans. Power Deliv.*, vol. 19, no. 2, pp. 894–896, 2004.
[15] A. Jovicic, M. Jereminov, L. Pileggi, and G. Hug, "An Equivalent Circuit Formulation for Power System State Estimation including PMUs," in *North American Power Symposium (NAPS)*, 2018.
[16] M. Jereminov, A. Pandey, H. A. Song, B. Hooi, C. Faloutsos, and L. Pileggi, "Linear Load Model for Robust Power System Analysis," in *IEEE PES Innovative Smart Grid Technologies Conference Europe*, 2017.
[17] M. Jereminov, D. M. Bromberg, A. Pandey, M. R. Wagner, and L. Pileggi, "Adjoint Power Flow Analysis for Evaluating Feasibility," *arXiv:1809.01569*, 2018.
[18] M. Jereminov, A. Pandey, and L. Pileggi, "Equivalent Circuit Formulation for Solving AC Optimal Power Flow," *IEEE Trans. Power Syst.*
[19] A. Jovicic, M. Jereminov, L. Pileggi, and G. Hug, "A Linear Formulation for Power System State Estimation including RTU and PMU Measurements."
[20] A. B. Birchfield, T. Xu, K. M. Gegner, K. S. Shetye, and T. J. Overbye, "Grid Structural Characteristics as Validation Criteria for Synthetic Networks," *IEEE Trans. Power Syst.*, vol. 32, no. 4, pp. 3258–3265, 2017.
[21] A. Pandey, M. Jereminov, M. R. Wagner, and L. Pileggi, "Robust convergence of Power Flow using Tx Stepping Method with Equivalent Circuit Formulation," in *Power Systems Computation Conference, Dublin*, 2018.
[22] S. W. Director and R. A. Rohrer, "The Generalized Adjoint Network and Network Sensitivities," *IEEE Trans. Circuit Theory*, vol. 16, no. 3, pp. 318–323, 1969.
[23] S. W. Director and R. A. Rohrer, "Automated Network Design-The Frequency-Domain Case," *IEEE Trans. Circuit Theory*, vol. CT-16, no. NO. 3, pp. 330–337, 1969.
[24] J. W. Bandler and M. A. El-Kady, "A new Method for computarized solution of power flow equations," *IEEE Trans. Power Apperatus Syst.*, vol. PAS-101, no. 1, pp. 1–10, 1982.
[25] L. A. F. M. Ferreira, "Tellegen's Theorem and Power Systems - New Load Flow equations, new solution method," *IEEE Trans. Circuits Syst.*, vol. 37, no. 4, pp. 519–526, 1990.
[26] S. Boyd and L. Vandenberghe, *Convex Optimization*. 2004.
[27] V. M. da Costa, N. Martins, and J. L. R. Pereira, "Developments in the newton raphson power flow formulation based on current injections," *IEEE Trans. Power Syst.*, vol. 14, no. 4, pp. 1320–1326, 1999.
[28] M. R. Wagner, A. Pandey, M. Jereminov, and L. Pileggi, "Robust Probabilistic Analysis of Transmission Power Systems based on Equivalent Circuit Formulation," pp. 0–5, 2018.
[29] M. E. O'Neill, "PCG: A Family of Simple Fast Space-Efficient Statistically Good Algorithms for Random Number Generation," 2014.
[30] C. Josz, S. Fliscounakis, J. Maeght, and P. Panciatici, "AC Power Flow Data in MATPOWER and QCQP Format: iTesla, RTE Snapshots, and PEGASE," pp. 1–7, 2016.